\documentclass[prl,twocolumn,Letters, superscriptaddress]{revtex4}
\usepackage{graphicx}
\usepackage{color}
\usepackage{amsmath,amssymb,amsopn}
\usepackage{bbm}
\usepackage{dsfont}

\def\bra#1{\mathinner{\langle{#1}|}}
\def\ket#1{\mathinner{|{#1}\rangle}}

\def\prjct#1{\mathinner{|{#1}\rangle}\!\!\mathinner{\langle{#1}|}}
\newcommand{\coh}[2]{\mathinner{|{#1}\rangle}\!\!\mathinner{\langle{#2}|}}
\newcommand{\braket}[2]{\langle #1  |#2\rangle}

\def\text#1{\textrm{#1}}
\def\id{\mathds{1}}
\def\E{\mathcal{E}}
\def\eq{\begin{equation}}
\def\eeq{\end{equation}}
\def\tr{\text{tr}}

\def\Q{\text{Q}}
\def\M{\text{M}}
\def\L{\mathcal{L}}
\def\norm#1{||#1||}
\def\N{\mathcal{N}}

\begin{document}

\date{\today}

\author{Pavel Sekatski}
\affiliation{Group of Applied Physics, University of Geneva, CH-1211 Geneva 4, Switzerland}
\affiliation{Institut f\"ur Theoretische Physik, Universit\"at Innsbruck, Technikerstr. 25, A-6020 Innsbruck, Austria}
\author{Nicolas Gisin}
\affiliation{Group of Applied Physics, University of Geneva, CH-1211 Geneva 4, Switzerland}
\author{Nicolas Sangouard}
\affiliation{Group of Applied Physics, University of Geneva, CH-1211 Geneva 4, Switzerland}

\title{How difficult it is to prove the quantumness of macroscropic states?}

\begin{abstract}
General wisdom tells us that if two quantum states are ``macroscopically distinguishable'' then their superposition should be hard to observe. We make this intuition  precise and general by quantifying the difficulty to observe the quantum nature of a superposition of two states that can be distinguished without microscopic accuracy. First, we quantify the distinguishability  of any given pair of quantum states with measurement devices lacking microscopic accuracy, i.e. measurements suffering from limited resolution or limited sensitivity. Next, we quantify the required stability that have to be fulfilled by any measurement setup able to distinguish their superposition from a mere mixture. Finally, by establishing a relationship between the stability requirement and the ``macroscopic distinguishability'' of the two superposed states, we demonstrate that indeed, the more distinguishable the states are, the more demanding are the stability requirements. \end{abstract}
\maketitle

\section{Introduction}

The predictions of quantum physics  are extremely well reproduced in experiments all over the world for almost a century. Even the most counter intuitive effects, such as entanglement and non-locality, were repeatedly confirmed. Those effects forbid a local interpretation of physical reality, that is at the core of classical physics. Once one accepts a radical departure form classical realism, it becomes quite puzzling that quantum physics is unnecessary when describing macroscopic phenomena, because there is no explicit quantum to classical transition mechanism within quantum theory itself. There are two ways to tackle this problem.
 
From one side, there are attempts to derive a new theory that reproduces quantum and classical physics as two asymptotic cases. Phenomenologically, those new theories can be seen as quantum theory supplemented with an explicit collapse mechanism, and might have a rich underlying physics, see \cite{Bassi13} for a recent review. 

On the other hand, there are attempts to mimic the arousal of classical reality from within quantum theory itself. Decoherence \cite{Zurek03} is usually given as a solution: a system unavoidably interacts with its environment, which measures its state and destroys the quantum correlations. It is argued that decoherence is more and more important when the ``size of the system'' increases, in such a way that in practice no quantum property can be observed in ``large systems''. Another possibility is to blame the measurement \cite{Mermin80, Peres02, Kofler07, Raeisi11}. Here, the central intuition is that revealing entanglement in a ``large system'' requires measurements with an extreme precision. The two approaches are closely related, since a decoherence channel can be seen as acting on the observable, spoiling the accuracy of the measurement.

A lot of examples have been presented in the literature confirming the general intuition that the quantum features of ``macroscopic'' states are very fragile with respect to experimental imperfections (uncontrolled interactions with the environment and imperfect measurements) and therefore extremely hard to observe. 

In this letter, we go beyond concrete examples and put this intuition in a quantitative form. Naturally, the first step is to elucidate the meaning of ``macroscopic quantum state'', that is not obvious. Indeed, measuring the size of a quantum state (either a superposition or an entangled state) is at the heart of many recent papers \cite{Dur02, Bjorn04, Korsbakken07,  Marquardt08, Lee11, Frowis12, Nimmrichter13, Sekatski13}, which result in various definitions, but they are not easily connected to decoherence/measurement inaccuracy. Here, we focus on a definition based on the observer point of view in which a system M containing a superposition of two states $\ket{A}_M$ and $\ket{D}_M$ is macroscopic if the two superposed states  can be distinguished with a very inaccurate measurement. 

Concretely, we consider Schr\"odinger cat like states $\ket{A}_\M\ket{\uparrow}_\Q + \ket{D}_\M \ket{\downarrow}_\Q$ where the system M is entangled with a microscopic two-level system Q. We characterize such a state by quantifying how well the components $\ket{A}_M$ and $\ket{D}_M$ can be distinguished with inaccurate measurements, the inaccuracy being either coarse-graining (limited resolution) or inefficiency (limited sensitivity), see below. In both cases, entanglement is reduced by experimental defects, namely by noise that we identify with a weak measurement \cite{Aharonov05} of the observable that is coarse-grained (for limited resolution), or by loss (for limited sensitivity). We prove that the amount of entanglement surviving the defects is upper bounded by an expression involving solely the probability to distinguish the two components with inaccurate measurements, see Eqs.~(\ref{bound noise}) and (\ref{bound loss}). This shows quantitatively that the observation of entanglement becomes progressively harder as the distinctness of the components increases. We also demonstrate that this holds for quantum superpositions instead of entangled states, i.e. the task to distinguish a superposition of two states from a statistical mixture becomes more and more difficult as the the distinctness of the superposed states increases. We conclude that observing the quantum nature of macro states, either described by quantum superpositions or by entangled states, requires an extreme control.

\section{Setting the problem}

Consider a bipartite entangled state
\eq\label{state}
\ket{\zeta}_{\M\Q}=\ket{A}_\M\ket{\uparrow}_\Q + \ket{D}_\M \ket{\downarrow}_\Q,
\eeq
where the system of interest M is in a superposition of states $\ket{A}$ and $\ket{D}$ that are potentially macroscopically distinct. The test qubit Q entangled with M contains two orthogonal components $\ket{\uparrow}$ and $\ket{\downarrow}$. It fixes what are the superposed components $\ket{A}$ and $\ket{D}$
\footnote{Since the work of Schr\"odinger it is customary to refer to the state \ref{state} as a macroscopic superposition, as if there was only the system M. We believe that the presence of Q is necessary. It fixes the superposed components $\ket{A}$ and $\ket{D}$ (up to a rotation), remark that $\ket{A}+\ket{D}=\ket{C}$ admits an infinite number of decompositions. Furthermore it allows one to talk about quantum correlations; in the absence of Q it is unclear how to certify that $\ket{A}$ and $\ket{D}$ are genuinely superposed (and not mixed) without knowing the Hilbert space where they belong to.}.

Let us take two orthogonal components $\braket{A}{D}=0$.
 Then, the states $\ket{A}$ and $\ket{D}$ can be distinguished with certainty in a single shot. However the measurement that allows one to do so might be very complicated, and more importantly this operational distinctness might be very fragile with respect to technical imperfections within the measurement device. We call a pair of states $\ket{A}$ and $\ket{D}$ macroscopically (or easily) distinguishable if this is not the case, i.e. if an imperfect measurement still allows one to distinguish them. 

To put it more formally let $\sigma$ be the parameter describing the imperfection of our measurement device (with $\sigma=0$ giving an ideal measurement). 
The probability to guess between the two states $\ket{A}$ and $\ket{D}$ in a single shot, labeled $P_g$ is then a function of $\sigma$. It generally drops when $ \sigma$ increases. The pair $(\sigma, P_g)$ characterizes the distinctness of states $\ket{A}$ and $\ket{D}$.

Easily distinguishable states are such that can be distinguished with high probability (high $P_g$) with an inaccurate measurement device (high $\sigma$) in a single shot. Below we  show that a superposition of easily distinguishable states is necessarily very fragile. We consider two aspects in which a measurement device can be imperfect, namely  a limited resolution (coarse-graining) or a limited sensitivity (probability to interact with the measured system). The intuition is that in each case the which-path information is easily extractable from their superposition either by noise or loss, which turns the superposition state into a statistical mixture.\\

\section{Distiguishability with inaccurate measurements}

\paragraph{Measurement with limited resolution} -- \-\ Consider an ideal measurement of an arbitrary operator $\hat X$. The probability (or probability density in the continuous case)  to observe an outcome x within the spectra of $\hat X$ is obtained from the operator $\delta(\hat X - x)$, i.e. for a state $\ket{S}$ it is $p_0^S(x) = \bra{S} \delta(\hat X-x)\ket{S}$. The effect of coarse-graining on $\hat X$ is to smear out the outcome distribution
\eq\label{coarse measurement}
p_\sigma^S(x)= \int g_\sigma(\lambda) p_0^S(x+\lambda) d\lambda = \bra{S} g_\sigma (\hat X-x)\ket{S}
\eeq
with a ``noise function'' $g_\sigma(\lambda)$ with zero mean and standard deviation $\sigma$. For concreteness we assume Gaussian noise in such a way that $g_\sigma$ is solely characterized by $\sigma$. 

The probability to correctly guess between two states $\ket{A}$ and $\ket{D}$ in a single shot  with a measurement device specified by $g_\sigma$ is given by
\eq\label{guess}
P_\sigma^{\hat X}[A,D]= \frac{1}{2}(1+D^{\hat X}_\sigma[A,D])
\eeq
with $D^{\hat X}_\sigma[A,D]=\frac{1}{2}\int|p^A_\sigma(x)-p^D_\sigma(x)|dx$  -- the trace distance between the outcome probability distributions corresponding to the two input states. Note also that to any fixed value of the guessing probability $P_g$ corresponds a value ${\cal R}^{\hat X}_{P_g}$
\eq
{\cal R}_{P_g}^{\hat X}[A,D] \equiv \max\{\sigma : P^{\hat X}_\sigma[A,D]\geq P_g \},
\eeq
that gives the worst possible measurement that allows one to distinguish the states with at least the required probability $P_g$. Both $P_\sigma^{\hat X}$ and ${\cal R}_{P_g}^{\hat X}$ characterize the distinctness of $\ket{A}$ and $\ket{D}$ with respect to $\hat X$.\\

\begin{figure}[ht!]
\includegraphics[width=6cm]{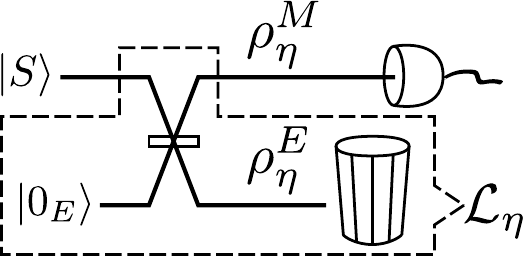}
\caption{Modelling a measurement device with limited sensitivity for photons.
} 
\label{fig1}
\end{figure}

\paragraph{Measurement with limited sensitivity} -- \-\ For a measurement device with limited sensitivity $\eta < 1$, there is a chance that the system or its part goes through without being detected. In other words, the interaction between the measured system and the measurement device is a probabilistic process. Such a process can be modelled by splitting the measured system into two parts and then sending only one part to an ideal measurement device (the other part is given to the environment and traced out). In the case of photons such a splitting is produced by a beamspliter with transmission $\eta$, and in the case of qubit ensembles it is produced by uncorrelated depolarisation channels. But such a probabilistic loss channel $\mathcal{L}_\eta$  can be defined in full generality, and it has to satisfy several conditions. Let $\rho_\eta^M =\mathcal{L}_\eta \big( \ket{S}\big)$be the partial state of the system after the interaction ($\rho_\eta^E$ is the partial state of the environment), cf \figurename{~\ref{fig1}}. Then (i) for a unit efficiency $\rho_{\eta=1}^M$ equals to the input state $\ket{S}$, (ii) for a zero efficiency $\rho_{\eta=0}^M=\ket{0_E}$ contains no information about the input state $\ket{S}$, (iii) the outputs are symmetric $\rho_\eta^M = \rho_{1-\eta}^E$\footnote{In fact one only requires symmetry up to a local unitary $U_M\rho_\eta^M U_M^\dag= U_E \rho_{1-\eta}^E U_E^\dag$. Note that it is also reasonable to assume: (iv) for a varying efficiency $\eta \in [0,1]$ the trajectory in the space of states $\rho_\eta^M$ is continuous.}.
The probability to distinguish $\ket{A}$ and $\ket{D}$ with a measurement device of efficiency $\eta$ is at most given by
\eq
P_\eta^\L[A,D] = \frac{1}{2}\big(1 + D_\eta^\L[A,D]\big)
\eeq
with the trace distance $D_\eta^\L[A,D]= \frac{1}{2}\tr | \mathcal{L}_\eta\big(\ket{A}\big) -\mathcal{L}_\eta\big(\ket{D}\big)|$. As before, by fixing the required guessing  probability $P_g$, one can extract the minimal efficiency ${\cal S}_{P_g}$ allowing to achieve it
\eq
{\cal S}_{P_g}^\L[A,D] \equiv \min\{\eta: P_\eta^\L[A,D]\geq P_g \}.
\eeq

\section{Fragility of entanglement}
We now analyze the fragility of entanglement contained in the state (\ref{state}) when the system M is sent through various decoherence channels $\E$. Specifically, for the states containing components that are easily distinguishable with limited resolution on $\hat X$, we first identify the noise channel that consists in applying a random unitary $e^{i \lambda \hat X}$ on the state (with the random variable $\lambda$) as a weak-measurement of $\hat X$ by the environment. It is then clear that this channel rapidly extracts the which-path information if the two components under consideration are easily distinguishable with $\hat X$ measurements. For the states containing components that are easily distinguishable with limited sensitivity, we use the symmetry of the outputs of the loss channel. In particular, we show that when such a superposition is sent through a loss channel, the environment can extract the which-path information rapidly, i.e. from a small fraction of the state, because the observer can do so. 

\begin{figure}[ht!]
\includegraphics[width=6cm]{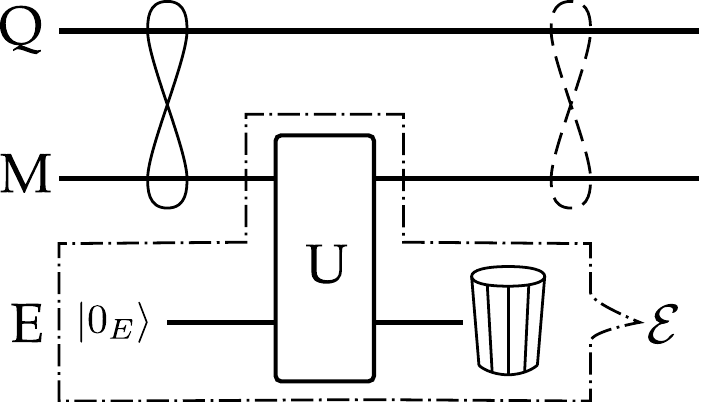}
\caption{\label{fig2} System-environment representation of the decoherence channel $\E$ acting on M.} 
\end{figure}

Quantitatively, consider the system-environment representation of some decoherence channel described by a global unitary $U$, as depicted in \figurename{~\ref{fig2}}. It can be shown that the negativity \cite{Horodecki09} of the state $\rho_f = \E\big(\ket{\zeta}_{\M\Q}\big)$ is upper bounded by the which-path information available to the environment after the interaction with the system (see Appendix)
\eq\label{bound}
2\N(\rho_f) \!\leq\!\sqrt{1\!-\!D(\rho_E^A, \rho_E^D)^2}\!\leq\!\sqrt{1\!-\!D_{\{E_m\}}(p^A_m,p^D_m)^2}
\eeq
where $D(\rho_E^A, \rho_E^D)$ is the trace distance between the partial states of the environment $\rho_E^S = \tr_M U\prjct{S,0_E}U^\dag$, and $D_{\{E_m\}}(p^A_m,p^D_m)$ is the trace distance between the distributions $\{p_m^A\}$ and $\{ p_m^D\}$ that are the probabilities of outcomes for the POVM $\{E_m\}$ and for the states $\rho_E^A$ and $\rho_E^D.$ Remark that $D(\rho_E^A, \rho_E^D)= \underset{\{E_m\}}{\max}\, D_{\{E_m\}}(p^A_m,p^D_m)$. Notice also that a strictly zero r.h.s in (\ref{bound}) not only ensures a zero negativity, but also proves that the state $\rho_f$ is separable (no PPT entanglement, see Appendix). We now apply this result to the superpositions of easily distinguishable states.\\

\paragraph{Noise and superpositions of states distinct with coarse-grained detectors} -- \-\ Consider the state (\ref{state}) with the functions $P_\sigma^{\hat X}[A,D]$ and ${\cal R}_{P_g}^{\hat X}[A,D]$ characterizing the distinctness of the components $\ket{A}$ and $\ket{D}$ under noisy measurements of $\hat X$. Further consider the following noise channel
\eq\label{stability channel}
\E_\Delta^{\hat X}( \rho)= \int e^{i \lambda \hat X }\, \rho \,e^{-i \lambda \hat X} f_\Delta(\lambda) d\lambda.
\eeq
It corresponds to diffusion of state $\rho$ in the directions complementary to $\hat X$ in the Hilbert space. This channel is characterized by the standard deviation $\Delta$ of the distribution $f_\Delta(\lambda)$, that we assume Gaussian.  So $\Delta$ describes the instability of the setup with respect to transformations generated by $\hat X$. 
We give three examples of such a channel below, but let us first show that it corresponds to a weak measurement of $\hat X$ (without postselection).
For this, we rewrite (\ref{stability channel}) in the system-environment representation depicted in \figurename{~\ref{fig2}}.
Let the environment be a particle in one dimension with the initial state $\ket{0_E}= \int \psi (q)\ket{q} dq = \int \tilde \psi(p) \ket{p} dp$. Then the channel (\ref{stability channel}) is equivalently given by
\eq\label{stability2}
\E_\Delta^{\hat X}( \rho) =\tr_E e^{i \hat p \hat X} \rho \otimes \prjct{0_E} e^{-i \hat p \hat X}
\eeq
for $|\tilde \psi(p)|^2 = f_\Delta(p)$. 
Eq. (\ref{stability2}) represents a weak measurement of the $\hat X$ observable performed by the environment:
 $e^{i \hat p \hat X}$ shifts the initial state of the environment $\psi(q)$ in the q-line by exactly $\hat X$.
So when reading its position,  the probability of having the outcome $q$ (corresponding to the projector $\prjct{q}$) equals $ \tr\, \id_M\! \otimes\! \prjct{q} e^{i \hat p \hat X} \rho \otimes \prjct{0_E} e^{-i \hat p \hat X} =\tr_M |\psi(q-\hat X)|^2\rho$. Identifying $|\psi(x)|^2 = g_\sigma(x)$ with $\sigma=\frac{1}{\Delta}$ the measurement of the position of the final state of environment corresponds to the coarse-grained measurement of $\hat X$ defined in (\ref{coarse measurement}). 
Combining (\ref{bound}) for $E_m= \prjct{q}$ and (\ref{guess}), one finds
\eq\label{bound noise}
\N\Big(\E_\Delta^{\hat X}\big(\ket{\zeta}_{\M\Q}\big)\Big)\leq  \sqrt{P_\frac{1}{\Delta}^{\hat X}[A,D](1-P_\frac{1}{\Delta}^{\hat X}[A,D])}.
\eeq
The entanglement remaining in the state $\ket{A}_\M\ket{\uparrow}_\Q + \ket{D}_\M \ket{\downarrow}_\Q$ after the channel $\E_\Delta^{\hat X}$ is upper bounded by an expression involving $P_\frac{1}{\Delta}^{\hat X}[A,D]$ -- the probability to guess between the two components $\ket{A}$ and $\ket{D}$ with a $\hat X$-measurement coarse-grained with $\frac{1}{\Delta}$ noise in a single shot. 
As an example, for the guessing probability $P_g= \frac{u^2}{u^2+4}$, the decoherence channel (\ref{stability channel}) with strength $\Delta=1/{\cal R}_{P_g}^{\hat X}$ reduces the entanglement by at least $u$.  In other words, when the two superposed components are well distinguishable with coarse-grained measurements (${\cal R}_{P_g}^{\hat X}\gg 1$), their superposition reduces to a statistical mixture even for tiny imperfection of the experimental setup ($\Delta= 1\big/ {\cal R}_{P_g}^{\hat X}\ll 1)$. Note that the channel $\E_\delta^{\hat X}$ can be equivalently seen as a decoherence affecting the state or as a lack of control on the measurement setup. In particular, this yields the following concrete results:

(I) A superposition of two states $\ket{A}$ and $\ket{D}$  distinct in energy $\hat X=\hat H_M$ is necessarily fragile with respect  to phase noise channel (\ref{stability channel}), which stands for a lack of control on time precision (or on the length of optical paths for photons).

(II) A superposition of states distinct with respect to the spatial position $\hat X=\hat x$ is necessarily very demanding on the precision of momenta measurement $\hat p$ to be revealed, since in this case the channel (\ref{stability channel}) corresponds to a coarse-graining of the momenta observable $\delta(\hat p) \to f_\Delta(\hat p)$. (Same applies for any quadrature measurements).

(III) A superposition of states distinct with respect to spin $\hat S_z$ is increasingly demanding on the control of the polar angle on the Bloch sphere, since the channel (\ref{stability channel}) for $\hat X =\hat S_z$ stands for random rotations around the $z$-axis.
\\

\paragraph{Loss and superpositions of states distinct with insensitive detectors} -- \-\ Consider the state (\ref{state}) with the functions $P_\eta^\L[A,D]$ and ${\cal S}_{P_g}^\L[A,D]$ characterizing the distinctness of the components $\ket{A}$ and $\ket{D}$ with insensitive detectors. The entanglement in this state after the loss channel $\mathcal{L}_\eta (\ket{\zeta}_{\M\Q})$ is upper bounded by the trace distance between environmental partial states  $\rho_\eta^E(\ket{A})$ and $\rho_\eta^E(\ket{D})$, see Eq.~(\ref{bound}). Using the symmetry property of the loss interaction outputs $\rho_\eta^E =\rho_{1-\eta}^M$ we find
\eq\label{bound loss}
\N\Big(\mathcal{L}_\eta \big(\ket{\zeta}_{\M\Q}\big)\Big) \leq  \sqrt{P_{1-\eta}^\L[A,D](1-P_{1-\eta}^\L[A,D])}.
\eeq
For example for $P_g = \frac{u^2}{u^2+4}$, any loss channel with transmission $\eta  \leq 1-{\cal S}_{P_g}^{\mathcal{L}}$ reduces the entanglement in the state $\ket{A}_\M\ket{\uparrow}_\Q + \ket{D}_\M \ket{\downarrow}_\Q$ by at least $u$.\\

\section{Concluding discussion}

We have shown in Eqs. (\ref{bound noise}) and (\ref{bound loss}) that for any state of the from (\ref{state}), the amount of entanglement that can be revealed in presence of experimental defects is limited  by the distinctness of components $\ket{A}$ and $\ket{D}$ that can be achieved with inaccurate measurements.  Several remarks naturally arise from this result.\\

\paragraph{Measure of Macroscopicity} -- \-\ The idea of looking at the distinctness between the components of a superposition was at the core of measures of macroscopicity presented in \cite{Korsbakken07} for insensitive measurements and in \cite{Sekatski13} for measurements with limited resolution. This definition seems very intuitive. Indeed, to distinguish a dead and an alive cat, one doesn't need measurements with precision at the level of a single atom contrary to the precision required for the observation of typical microscopic properties. 
Along the same lines, we note that the maximal tolerable inaccuracy ${\cal R}_{P_g}^{\bar X}$ or ${\cal S}_{P_g}^\mathcal{L}$ itself (or any monotonous function of these parameters) can be used as a measure of the size of a superposition state \footnote{ For example, it is appropriate to calibrate the size $N$ with a family of superposition states $\ket{\Psi(N)}$ with a naturally defined size, as it was done with superpositions of Fock states in \cite{Sekatski13}}.\\

\paragraph{Certifiability} -- \-\ Instead of talking about entanglement one could drop the qubit from the state (\ref{state}) and formulate the problem differently:
Is it possible  after the decoherence channel $\E$ (or $\L$) to certify that the state was prepared in a superposition 
\eq \label{Certifiability state}
\ket{\xi_S}= \ket{A}+\ket{D}
\eeq
 and not a mixture $\ket{\xi_M}= \prjct{A}+ \prjct{D}$? This problem is closely related to the notion of certifiablilty introduced in \cite{Frowis13}, where the authors showed for the case of qubit ensembles and uncorrelated depolarizing noise, the superpositions of macroscopically distinct states are incertifiable. Our findings allow us to draw similar conclusions. By replacing the negativity $\N$ by the trace distance $D[\E\big(\ket{\xi_S}\big),\E\big( \ket{\xi_M} \big)]$ in inequalities (\ref{bound noise}) and (\ref{bound loss}) (as follows from equations (16-18) in the Appendix), we conclude that to certify the superposition of easily distinguishable states becomes more and more difficult as the distinguishability increases.\\

\paragraph{Observing entanglement with coarse-grained measurements} -- \-\ In Ref.~\cite{Wang13}, it has been conjectured that in order to ``{\it detect quantum effects such that superposition or entanglement in macroscopic systems either the outcome precision or the control precision of the measurements has to increase with the system size}''. The conjecture was illustrated  with an example involving two coherent states $\ket{A} = \ket{\alpha}$ and $\ket{D}= \ket{-\alpha}$.  By increasing $\alpha$ one can make these states distinguishable even with low outcome precision in the computational basis. However, to reveal a quantum feature it is necessary to also perform a measurement in another basis. To do so, one can apply a non-linear control transformation $U_\theta= e^{i \theta \hat N^2}$ on the state, before measuring again in the computational basis. The authors showed that when $\alpha$ increases, the constraint on the precision of the angle $\theta$ increases as well. Our results allow us to generalize this result. Consider a measurement of $\hat{X}$ (given by $f(\hat X)$) preceded by a control transformation $U$ on a state $\E_\Delta^{\hat X} \big(\rho\big )$. Since $e^{i \lambda  \hat X}$ commutes with $\hat X$ one has 
\eq \tr \, U^\dag  f(\hat X) \, U \E_\Delta^{\hat X}\big( \rho \big) = \tr\,  f(\hat X) \, \int d \lambda p(\lambda) U_\lambda \, \rho \, U_\lambda^\dag, 
\eeq
with $U_\lambda = e^{-i \lambda \hat X} U e^{i \lambda \hat X}$. Therefore, in this context, the channel $\E_\Delta^{\hat X}$ acting on the state can be interpreted as a lack of the control precision (on U). Reciprocally any lack of control precision generated by $\hat X$ can be seen as a noise channel operating on the state and equivalently as a weak measurement of $\hat X$ by the environment. This shows in full generality that the demand on the control precision increases with the distinctness of the components $\ket{A}$ and $\ket{D}$ as suggested by (\ref{bound noise}).\\

\paragraph{Effective classical-to-quantum transition} -- \-\ Finally, let us emphasise that our results can be seen as an effective bound between the classical and the quantum domains. If the macroscopic quantum states are defined as those containing a superposition of components that can be distinguished with imperfect measurements, then their quantum nature is very difficult to observe.\\

\paragraph{Acknowledgments} We thank Wolfgang D\"ur, Florian Fr\"owis, Bruno Sanguinetti and Rob Thew for helpful discussions and comments. We acknowledge support by the ERC-MEC, the Suiss NSF projects "Large Entanglement in Crystals" and "Early PostDoc.Mobility", and the Austrian Science Fund FWF P24273-N16.

\section{Appendix}

Here we derive the bound (\ref{bound}) which describes how the which path information available to the environment after decoherence affects the entanglement in the state (\ref{state}). Consider a decoherence channel $\E$ acting on the system M.
We explicit the matrix structure of Q with $\ket{\uparrow}_\Q= \genfrac{(}{)}{0 pt}{1}{1}{0}$  and $\ket{\downarrow}_\Q= \genfrac{(}{)}{0 pt}{1}{0}{1}$, the state $\E\big(\ket{A}_\M\ket{\uparrow}_\Q + \ket{D}_\M \ket{\downarrow}_\Q\big)$ after the decoherence channel reads
\eq\label{final state}
\rho_f =\left(\begin{array}{cc}
\E\big(\prjct{A}\big) & \E\big(\coh{A}{D}\big)\\
\E\big(\coh{D}{A} \big) & \E\big(\prjct{D} \big)
\end{array}\right).
\eeq 
Using the system-environment representation of the decoherence channel depicted in \figurename{~\ref{fig2}}, the state reads
\eq
\tr_E \left(\begin{array}{cc}
U \prjct{A,0_E}U^\dag & U\coh{A,0_E}{D,0_E} U^\dag\\
U \coh{D,0_E}{A,0_E} U^\dag & U \prjct{D,0_E} U^\dag
\end{array}\right).
\eeq 
Consider a POVM defined on the environment with $\id_E= \sum_m E_m^\dag E_m$, insert it inside the trace in the above expression and bring the Kraus operators $E_m$ ($E_m^\dag$) on the left (right) side. Any component $E_m U \ket{A,0_E} = \sqrt{p_m^A} \ket{v_m^A}$ (and $E_m U \ket{D,0_E} = \sqrt{p_m^D} \ket{v_m^D}$ respectively), where $ \ket{v_m^{A(D)}}$ is a normalized state of the joint system $\M\otimes \text{E}$, and $p_m^{A(D)}=\norm{E_m U \ket{A(D),0_E}}$ is the probability of the POVM outcome ``m'' for the state $\ket{A} (\ket{D})$ in M. Accordingly the state $\rho_f$ is rewritten as a sum of terms
\eq\label{decomposition}
\sum_m \tr_E \left(\begin{array}{rr}
p_m^A \prjct{v_m^A} & \sqrt{p_m^A p_m^D} \coh{v_m^A}{v_m^D}\\
\sqrt{p_m^A p_m^D} \coh{v_m^D}{v_m^A} & p_m^D \prjct{v_m^D}
\end{array}\right).
\eeq 
This expression is useful to bound the entanglement in the state (\ref{final state}) (we use the negativity $\N$ as entanglement measure \cite{Horodecki09}). The negativity of a fixed m term in (\ref{decomposition}) is upper bounded by $\sqrt{p_m^A p_m^D}$ (achieved for $\braket{v_m^A}{v_m^D}=0$), moreover $N$ is  convex and non-increasing under partial trace. This leads to an upper-bound on the negativity of the state $\rho_f$
\eq
\N(\rho_f)\leq \frac{1}{2}\sum_m \sqrt{p_m^A p_m^D}.
\eeq  
 The term on the r.h.s. is the fidelity $F\big(\{p_m^A\},\{p_m^D\}\big)$ between the probability distributions $\{p_m^A\}$ and $\{p_m^D\}$ provided by the POVM $\{ E_m \}$. The fidelity is always  upper bounded by an expression involving the trace distance $F\leq \sqrt{1-D^2}$ \cite{Nielsen00}, yielding
\eq\label{entanglement bound}
2\N(\rho_f)\! \leq F\big(\{p_m^A\},\{p_m^D\}\big) \leq\! \sqrt{1\!-\!D\big( \{p_m^A \}, \{p_m^D\} \big)^2}.
\eeq
Notice that when the r.h.s is strictly zero, not only the negativity drops to zero, but also the state cannot be PPT entangled. Indeed $F=0$ implies $p_m^A p_m^D = 0$ for all m, and thus $\rho_f$ is a convex sum of separable states. 

Finally the negativity is also upper bounded by statistical distances between the partial states of the environment 
\eq 
2\N(\rho_f)\leq F(\rho_E^A, \rho_E^D) \leq \sqrt{1-D(\rho_E^A, \rho_E^D)^2}
\eeq
with $\rho_E^{A(D)}=\tr_M U\prjct{A(D),0_E} U^\dag$), since both the fidelity and trace distance can be defined from the optimizations over all possible POVMs $F(\rho_E^A, \rho_E^D)= \underset{\{E_m\}}{\text{min}}F\big(\{p_m^A\},\{p_m^D\}\big)$ and $D(\rho_E^A, \rho_E^D)= \underset{\{E_m\}}{\text{max}}D\big(\{p_m^A\},\{p_m^D\}\big)$ \cite{Nielsen00}.
\\

\end{document}